\documentclass[aps,pre,twocolumn,showpacs, amsmath,amssymb,amsfonts,floatfix,superscriptaddress,
groupedaddress]{revtex4-1}

\usepackage{graphicx}
\usepackage{dcolumn}
\usepackage{bm}
\usepackage{hyperref}
\usepackage{amsmath}
\usepackage{amssymb}
\usepackage{amsfonts}
\usepackage{sidecap}
\usepackage{float}
\usepackage{parskip}
\usepackage{grffile}

\usepackage{color}
\usepackage{hhline}
\makeatletter
\def\@eqnnum{{\normalsize \normalcolor (\theequation)}}
\makeatother
\begin{document}
	\title{Inter-layer adaptation induced explosive synchronization in multiplex networks}
	\author{Anil Kumar$^{1}$, Ajay Deep Kachhvah$^{1}$, Sarika Jalan$^{1,2}$} 
	\email{Corresponding Author:sarikajalan9@gmail.com} 
	\affiliation{$^1$Complex Systems Lab, Discipline of Physics, Indian Institute of Technology Indore, Khandwa Road, Simrol, Indore 453552}
	\affiliation{$^2$Centre for Bio-Science and Bio-Medical Engineering, Indian Institute of Technology Indore, Khandwa Road, Simrol, Indore-453552 }
	
	\date{\today}
	
	\begin{abstract}
		It is known that intra-layer adaptive coupling among connected oscillators instigates explosive synchronization (ES) in multilayer networks. Taking an altogether different cue in the present work, we consider inter-layer adaptive coupling in a multiplex network of phase oscillators and show that the scheme gives rise to ES with an associated hysteresis irrespective of the network architecture of individual layers. The hysteresis is shaped by the inter-layer coupling strength and the frequency mismatch between the mirror nodes. We provide rigorous mean-field analytical treatment for the measure of global coherence and manifest they are in a good match with respective numerical assessments. Moreover, the analytical predictions provide a complete insight into how adaptive multiplexing suppresses the formation of a giant cluster, eventually giving birth to ES. The study will help in spotlighting the role of multiplexing in the emergence of ES in real-world systems represented by multilayer architecture. Particularly, it is relevant to those systems which have limitations towards change in intra-layer coupling strength.
	\end{abstract}
	
	\pacs{89.75.Hc,05.45.Xt}
	
	\maketitle
	
	\section{Introduction} Recently, an irreversible synchronization process, called explosive synchronization \cite{Gardenes2011,Boccaletti2016}, in which a group of incoherent dynamical units is abruptly set in collective coherent motion, has drawn much attention of the researchers \cite{Raisa2019}. The abnormal hypersensitivity of ES can be perilous in many physical and biological circumstances such as abrupt cascading failure of the power-grid \cite{Buldyrev2010}, breakdown of the internet due to intermittent congestion \cite{Huberman1997}, abrupt attack of epileptic seizures in the human brain \cite{Adhikari2013} and abrupt episodes of chronic pain in the Fibromyalgia human brain\cite{Lee2018}, to name a few. An anesthetic-induced transition to unconsciousness \cite{Kim2016, Kim2017} and bistability in Cdc2-cyclin B in embryonic cell cycle \cite{Pomerening2003} are a few other instances of an abrupt transition. 
	
	The occurrence of ES transition has also been demonstrated experimentally \cite{Leyva2012, Motter2013, Kumar2015}. The emergence of ES is shown to be rooted in inertia \cite{Tanaka1997} and a microscopic correlation between frequency and degree or coupling strength of the networked phase oscillators \cite{Gardenes2011, Zhang2013}. Recently, Zhang {\em et al.}\cite{Zhang2015} showed that a fraction of adaptively coupled phase oscillators gives birth to ES. Adaptation is an inherent feature in the construction of many complex systems, for instance, adaptation in neuronal synchronization of brain regions apropos learning or memory process \cite{Zhou2006,Gutierrez2011}.

	\begin{figure}[t]
		\centering
		\centerline{\includegraphics[width=1\columnwidth]{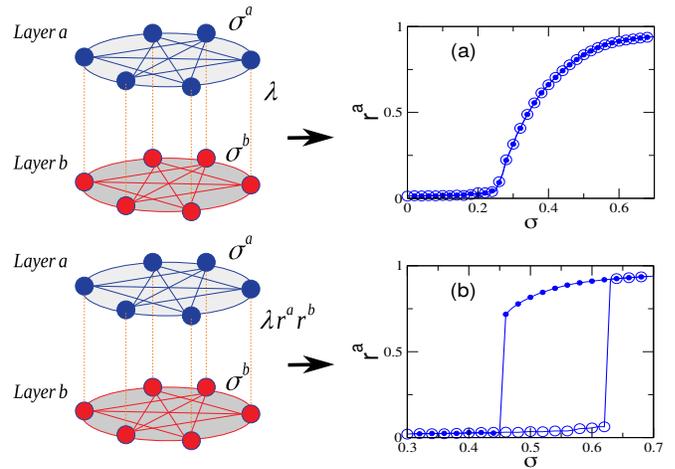}}
		\caption{(color online). In a multiplex network, (a) static inter-layer coupling ($\lambda$) leads to continuous phase transition while (b) adaptive inter-layer coupling ($\lambda r^ar^b$) leads to ES. 
			Two different type of transitions ($r^a$ vs $\sigma$) in (a) and (b) are shown for two globally connected networks, each of size $N=5000$ having $\lambda$=0.5. Circle and bullet with solid lines respectively correspond to forward and backward continuation of $\sigma$, unless stated elsewhere, throughout the paper.} 
		\label{multiplexing_effect}
	\end{figure}
	
	In many complex systems, the same set of nodes may have different types of connections among them, having each type influencing the functionality of other types. 
	Hence, an isolated network is an unfit candidate to model such systems. Such systems can be precisely framed by a multiplex network, where different layers denoting different dynamical processes are interconnected by the same set of nodes denoting interacting dynamical units \cite{DeDomenico2013,Sorrentino2012,Boccaletti2014,Jalan2014,Sevilla2016,DelGenio2016,Kumar2017,Pitsik2018,Leyva2018}. For instance, social networks, neuronal networks, and transport systems in which individuals, neurons, and cities have different types of connections among them forming different layers \cite{Kivela2014}. The multiplex framework has been remarkably successful in providing insights into the dynamical behavior of various processes such as percolation \cite{Osat2017}, epidemic \cite{Sahneh2013, DeDomenico2016b}, voter model \cite{Diakonova2016}, etc., taking place in a combination within a group, community or population.

	Recently, the studies apropos the occurrence of ES have been extended to multilayer networks by employing different methodologies, for instance, adaptive coupling \cite{Zhang2015}, intertwined coupling \cite{Nicosia2017}, inertia \cite{Ajaydeep2017}, delayed coupling \cite{Ajaydeep2019}, inhibitory coupling \cite{Vasundhara2019}, frequency mismatch \cite{Anil2019}. A few recent studies on multilayer networks have adopted the adaptive coupling dynamics proposed by Zhang {\em et al.} \cite{Danziger2016, Khanra2018, Danziger2019} at the heart of their models. 
	It is reported that the occurrence of ES in a multilayer network is exhibited by a fraction of nodes adaptively coupled via local order parameter, within the multiplexed layers, having virtual inter-layer links \cite{Zhang2015}. In the current work, taking an altogether different cue, we propose an approach in which inter-layer links of a multiplex network are adaptively coupled through global order parameters of the interacting layers, which brings about ES with an associated hysteresis. 
	Our model is sensitive to frequency mismatch between the interconnected nodes which, along with inter-layer coupling strength, determines the size of emergent hysteresis. Furthermore, the proposed model is robust against a variety of network topology in giving birth to ES. We corroborate our findings by providing rigorous mean-field analysis and a good match between the analytical prediction of the order parameter and its numerical evaluation.

	\section{Model and Technique}
	We consider a multiplex network consisting of two layers, each one having $N$ nodes represented by Kuramoto oscillators. Each node in a layer is adaptively linked with its counterpart in another layer through multiplexing strength which is a function of a measure of global coherence of the interacting layers. The time-evolution of phases of the nodes in the multiplexed layers {\it a} and {\it b} is ruled by \cite{Kuramoto1984}:
	\begin{subequations}
		\begin{align}
		\dot{\theta^{a}_i} = \omega_i^{a} + \frac{\sigma^a}{N} \sum_{j=1}^N A_{ij}^{a} \sin(\theta_j^{a}-\theta_i^{a})+
		\lambda r^a r^b\sin(\theta_i^{b}-\theta_i^{a}), \tag{1-a}\label{eq1} \\
		\dot{\theta_i^{b}} = \omega_i^{b} + \frac{\sigma^b}{N} \sum_{j=1}^N A_{ij}^{b} \sin(\theta_j^{b}-\theta_i^{b})+
		\lambda r^a r^b \sin(\theta_i^{a}-\theta_i^{b}), \tag{1-b}\label{eq2}
		\end{align}\label{model} 
	\end{subequations}
	where $i=1,...,N$. $\theta_i^{a(b)}$ and $\omega_i^{a(b)}$ represent phase and natural frequency of $i^{th}$ node in layer $a$($b$), respectively. $\sigma^{a(b)}$ denotes intra-layer coupling strength among the nodes of layer $a(b)$, here $\sigma^a = \sigma^{b} = \sigma$. The connectivity of the nodes in the multiplexed layers are encoded into a set of adjacency matrices ${\bf A}=\lbrace A^{a}, A^{b}\rbrace$, where $A^{a(b)}_{ij}$=1 if nodes $i^{a(b)}$ and $j^{a(b)}$ are connected, and $A^{a(b)}_{ij}$=0 otherwise. 
	$\lambda$ represents the inter-layer coupling strength or multiplexing strength. The multiplexing strength between the two layers is adaptively controlled by the product of global order-parameters ($r^a r^b$), a measure of degree of phase coherence among the nodes given by
	\begin{equation}
	r^{a(b)} e^{i\psi^{a(b)}} = \frac {1}{N} \sum_{j=1}^N  e^{i \theta^{a(b)}_{j}}
	\label{eq_r_a}
	\end{equation}
	where $0 \le r^{a(b)} \le 1$. $r^{a(b)}=0$ corresponds to a random distribution of the nodes over a unit circle, whereas $r^{a(b)}=1$ corresponds to exact phase synchronization. Eq.\ref{model} is numerically solved using RK4 method with time-step $dt=0.01$. 
	To determine phase coherence among nodes of layer $a$, time average of $r^a$ is taken for $10^5$ steps after neglecting initial $10^5$ steps. Initial phases and natural frequencies of the nodes in layers $a$ and $b$ are drawn from a uniform random distribution in the range $-\pi \le \theta_{i}^{a(b)} \le \pi$ and $-\gamma \le \omega_{i}^{a(b)} \le \gamma$, respectively. In this work we take $\gamma=0.5$.

	\section{Results}
	To investigate the effect of adaptive coupling between all the pairs of mirror nodes, we consider a multiplex network of two globally coupled networks, otherwise mentioned elsewhere. Furthermore, oscillators in the two layers have identical natural frequency distribution but in general, $\omega_i^a \neq \omega_i^b$.  
	Fig.\ref{multiplexing_effect} depicts behavior of $r^a-\sigma$ profile for the considered multiplex network. In the absence of adaptive multiplexing, the usual static $\lambda$ gives rise to a continuous phase transition in layer $a$, (Fig.\ref{multiplexing_effect}(a)). However, it unfolds that the presence of adaptive multiplexing strikingly leads to a discontinuous phase transition (ES) accompanied by a hysteresis, (Fig.\ref{multiplexing_effect}(b)). 
	
\begin{figure}[t]
		\centering
		\centerline{\includegraphics[width=1\columnwidth]{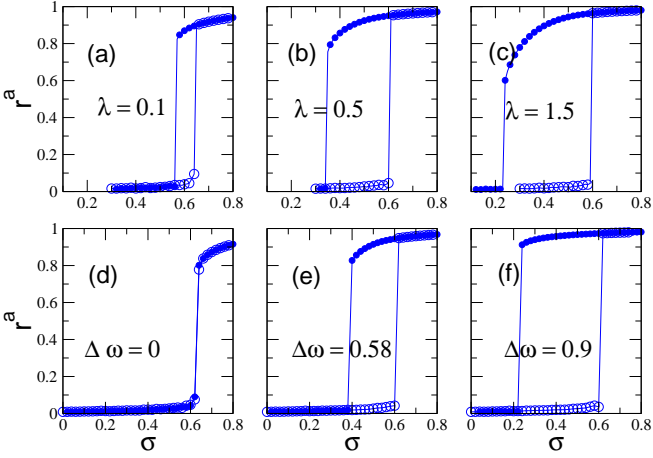}}
		\caption{(color online). (a) $r^a$ vs $\sigma$ for different multiplexing strength. (b) $r^a$ vs $\sigma$ for different $\Delta \omega$ values between the mirror nodes. In (a) $\Delta \omega = 0.58$, in (b) $\lambda =0.4$, while $N=10000$ in all the figures. 
		} 
		\label{Hysterisis_size}
	\end{figure}

\begin{figure}[t]
		\centering
		\centerline{\includegraphics[width=0.7\columnwidth]{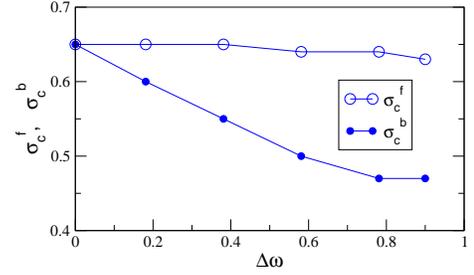}}
		\caption{(color online). $\sigma_c^f$ and $\sigma_c^b$ as a function of $\Delta \omega$. Here $N=10000$ and $\lambda=0.2$. 
		} 
		\label{forward_and_back_coupling_vs_omega}
	\end{figure}

	{\bf Factors determining hysteresis width:} 
	Fig.\ref{Hysterisis_size}, further elaborates on how inter-layer coupling strength ($\lambda$) and frequency mismatch between the mirror nodes affects the transition to synchronization in layer $a$. It turns out that an increase in $\lambda$ widens the hysteresis width (Fig.\ref{Hysterisis_size}(a-c)) associated with the emergent ES. Similarly, for a given $\lambda$, an increase in frequency mismatch, $\Delta\omega$, between mirror nodes also widens the hysteresis width (Fig.\ref{Hysterisis_size}(d-e)). Only the backward critical coupling ($\sigma_c^b$) manifests a decrease with an increase in $\lambda$ as well as $\Delta\omega$ while the forward critical coupling ($\sigma_c^f$) remains almost the same. 
	To measure the strength of frequency mismatch between the mirror nodes, we consider $\Delta \omega = 1- \frac{1}{2\sum_{i=1}^N |\omega_i^a|}\sum_{i=1}^N |(\omega_i^a + \omega_i^b)|$; $ 0\le\Delta \omega \le 1$. Here, $\Delta \omega =0$ if $\omega_i^a =\omega_i^b$ and $\Delta \omega =1$ if $\omega_i^a = - \omega_i^b$.  
	To obtain a desired value of $\Delta \omega$, starting with $\Delta \omega =0$, two pairs of mirror nodes are chosen randomly and their natural frequencies are swapped within the layers. After each swapping, $\Delta \omega$ is calculated and the change is accepted if the newer value of $\Delta \omega$ is closer to the desired value, otherwise, the change is discarded. This process is repeated until we get a desired value of $\Delta \omega$. Fig. \ref{forward_and_back_coupling_vs_omega} corroborates that an increase in  $\Delta \omega$ results in widening of hysteresis width ($\sigma^f_c-\sigma^b_c$). Hence, appropriate choices for both $\lambda$ and $\Delta\omega$ determine the threshold of explosive transition to desynchronization. A comparison between Fig. \ref{Hysterisis_size} (f) and Fig. \ref{forward_and_back_coupling_vs_omega} shows that the increase in $\lambda^f_c-\lambda^b_c$ is larger if $\lambda$ is larger. At $\Delta\omega=0.9$, Fig. \ref{Hysterisis_size} (f) shows that $\lambda^f_c-\lambda^b_c \approx 0.4$, while in Fig. \ref{forward_and_back_coupling_vs_omega} it is around $0.2$. 

\begin{figure}[t]
		\centering
		\centerline{\includegraphics[width=\columnwidth]{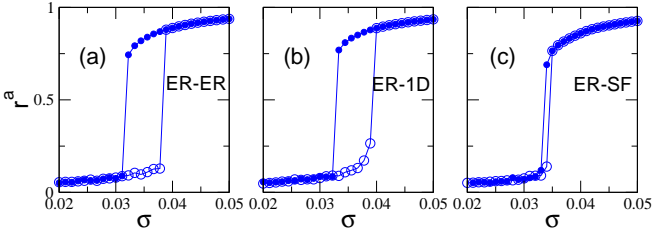}}
		\caption{(color online). (a-c) $r^a$ vs $\sigma$, showing robustness of ES against change in the network topology.
			Here $N=1000$, $\lambda=3$, while average connectivity ($\langle k ^{a(b)}\rangle $) of the two layers are $\langle k ^{a}\rangle $ = $\langle k ^{b}\rangle =16$. 
		} 
		\label{topology_effect}
	\end{figure}

	{\bf Robustness of ES against network topology:}
	The employed technique is robust against a variety of topologies such as regular ring network, Erd\"os-R\'enyi (ER) random network \cite{Erdos1959} and scale-free (SF) network \cite{barabasi1999} chosen for the individual layers.
	For numerical simulations, we replace intra-layer coupling $\sigma^{a(b)}/N$ in Eq. \ref{model} by $\sigma^{a(b)}$ \cite{Arenas2008}. 
	Fig. \ref{topology_effect} (a-c) shows that a multiplex network comprising either ER-ER network, ER-regular ring network or ER-SF network does exhibit ES transition with hysteresis as depicted by globally connected layers in Fig.~\ref{multiplexing_effect} and Fig.~\ref{Hysterisis_size}. However, hysteresis size may depend on the network topology as found in the case of ER-SF networks in Fig. \ref{topology_effect} (c).

\begin{figure}[t]
		\centering
		\centerline{\includegraphics[width=\columnwidth]{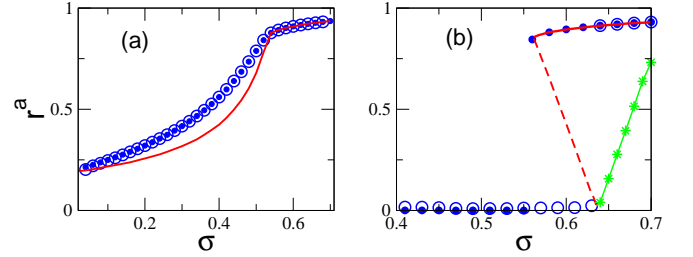}}
		\caption{(color online)
			(a) $r^a$  vs $\sigma$ in case of without adaptive coupling, and (b) with adaptive coupling. 
			Solid line, dashed line, and solid line with stars correspond to solutions of Eq. \ref{r_lay_a}. 
			Here $\lambda=0.1$, $\Delta \omega =1$, $N=10000$.} 
		\label{analytic_numeric}    
	\end{figure}

	{\bf Mean-field analysis when $\Delta \omega =1$:}
	To understand the mechanism behind the origin of ES, we analytically derive $r^{a(b)}$ for the synchronized state. The natural frequency of each node $i$ in layer $a$ is drawn from a uniform random distribution from interval $\omega_i^{a}= -0.5 +(i-1)/(N-1)$, $i=1, 2,\hdots,N$. Now, we take $\Delta \omega =1$, i.e., $\omega_i^{b}= - \omega_i^{a}$, Eq.\ref{model} can then be rewritten in terms of order parameter (Eq. \ref{eq_r_a}) as:
	\begin{subequations}
		\begin{align}
		\dot{\theta^{a}_i} = \omega_i^{a} + \sigma r^a \sin(\psi^a-\theta_i^{a})+
		\lambda A \sin(\theta_i^{b}-\theta_i^{a}), \tag{3-a}\label{eq3a} \\
		\dot{\theta^{b}_i} = -\omega_i^{a} + \sigma r^b \sin(\psi^b-\theta_i^{b})+
		\lambda A \sin(\theta_i^{a}-\theta_i^{b}), \tag{3-b}\label{eq3b}
		\end{align}\label{model_interms_of_r}
	\end{subequations}
	where $i=1,\hdots,N$, and $A=r^a r^b$. 
	In the synchronous state $\dot \theta_i^{a(b)} = \bar \omega^a$ \cite{Zhang2015,Gardenes2011}, where $\bar \omega^a = (1/N)\sum_{i=1}^N \omega_i^a =0$, hence the synchronous state is a fixed point, i.e., $\dot \theta_i^{a(b)} = 0$.
	Eq. \ref{model_interms_of_r}(a,b) suggests that a fixed point is possible when the values of phases are such that $\theta_i^a = - \theta_i^b$ and $\psi^a = - \psi^b$. By substituting the values of phases in Eq.\ref{eq_r_a}, one can observe that $r^a = r^b = r$. On assuming $\psi^a = \psi^b =0$ \cite{Strogatz2000}, Eq. \ref{model_interms_of_r}(a) results into
	\begin{equation}
	\omega_i^{a} - \sigma r \sin(\theta_i^{a})-
	\lambda A \sin(2 \theta_i^{a}) = 0. 
	\label{model_interms_of_r^a}
	\end{equation}
	After substituting $\sin(2 \theta_i^{a}) = \pm 2 sin (\theta_i^{a}) \sqrt{1-sin^2(\theta_i^{a})}$, Eq. \ref{model_interms_of_r^a} results in a fourth order polynomial $x_i^4 + p_i x_i^3 + q_i x_i^2+ r'_i x_i+ s_i =0 $, where $ x_i= \sin(\theta_i^a), p_i=0, q_i= (\sigma^2 r^2 / 4 \lambda ^2 A^2)-1, r'_i = -2\omega_i^a \sigma r / 4 \lambda ^2 A^2$, and $s_i =  (\omega_i^a)^2 / 4 \lambda ^2 A^2$. The roots of the polynomial are \cite{roots} 
	\begin{equation}
	\begin{split}
	x_{i_{1,2}} &=   \frac{R_i \pm D_i}{2},\\
	x_{i_{3,4}} &=  -\frac{(R_i \pm E_i)}{2},
	\end{split} 
	\label{roots_poly_4th_order}
	\end{equation}
	where 
	\begin{equation}
	\begin{split}
	R_i= \sqrt{z_i -q_i}, \\
	D_i=
	\begin{cases} 
	\sqrt{-R_i^2 -2q_i -2r'_i/R_i} & if R_i \neq 0 \\
	\sqrt{-2q_i +2\sqrt{z_i^2 -4s_i}} & if R_i =0, \\
	\end{cases} \\
	E_i=
	\begin{cases} 
	\sqrt{-R_i^2 -2q_i +2r'_i/R_i} & if R_i \neq 0 \\
	\sqrt{-2q_i -2\sqrt{z_i^2 -4s_i}} & if R_i =0. \\
	\end{cases}
	\end{split} 
	\label{coefficient_poly_4th_order}
	\end{equation}
	Here $z_i$ is a real root of the polynomial $z_i^3 -q_i z_i^2 -4 s_i z_i +4q_is_i - {r'_i}^2=0$ \cite{roots}. 
	Out of the four roots, a root with physically accepted solution is selected, as follows: Eq. \ref{model_interms_of_r^a} suggests that for any two nodes $i, j$ if $\omega_j^a = - \omega_i^a$, we get $\theta_j^a = - \theta_i^a$. Therefore, for the synchronous solution, we consider only
those roots of the polynomial which satisfy this condition for phases.
	Eq. \ref{coefficient_poly_4th_order} implies that if $\omega_j^a = -\omega_i^a$, then $R_i=R_j$ as $q_i$ is independent of $\omega_i^a$ and the third order polynomial also does not depend on the sign of $\omega_i^a$, hence, $z_i$ is also independent of sign of $\omega_i^a$. Moreover, for any $\omega_i^a \neq 0,\ R_i \neq 0$ as for $R_i = 0$ or $z_i =q_i$, the third order polynomial would imply $r_i' = 0$, which is not possible unless $\lambda$ is infinite.  
	Eq. \ref{coefficient_poly_4th_order} also implies that $D_i=E_j$, therefore $x_{i_1} = -x_{i_3}$ and $x_{i_2} = -x_{i_4}$. Thus, a possible pair of the roots are either $x_{i_1},x_{i_3}$ or $x_{i_2},x_{i_4}$. We find that $r$ values corresponding to $x_{i_1}$ and $x_{i_3}$ do not match with the numerical simulations (see star with solid line in Fig. \ref{analytic_numeric} (b)), hence the synchronous state corresponds to $x_{i_2}$ and $x_{i_4}$. It is quite apparent that for $\omega_i^a =0$,  $x_{i_2} = x_{i_4} =0$. 
	After plugging the values of phases from Eq. \ref{roots_poly_4th_order} into Eq. \ref{eq_r_a}, Eq. \ref{eq_r_a} can be rewritten as sum of the contributions from the locked and the drifting oscillators as: 
	\begin{equation}
	r^a =\frac {1}{N}  \sum_{locked}  \cos{ (\theta^{a}_{j}- \psi^a)} + \sum_{drifting} \cos{ (\theta^{a}_{j}- \psi^a)}
	\label{r_lay_a}
	\end{equation}
	For a given $\sigma$ and $\lambda$, natural frequency of the locked oscillator must satisfy the following relation 
	\begin{equation}
	|\omega_i^a| \le max|\sigma r \sin(\theta_i^a) + \lambda A \sin(2 \theta_i^a)|.
	\label{omega_constrain_to_synch}
	\end{equation} 
	If $f(\theta^a)$ be the RHS of Eq. \ref{omega_constrain_to_synch}, then
	$f(\theta^a)$ has extrema at $\cos{{(\theta_i^a)}^*_{\pm}} = \frac{-\sigma r}{8 \lambda A} \pm \sqrt{ \frac{\sigma^2 r^2}{64 \lambda^2 A^2} + 0.5}$ obtained from the roots of $d f/d \theta^a =0$. 
	Substituting the condition for extrema in Eq. \ref{omega_constrain_to_synch}, we get
	\begin{equation}
	|\omega_i^a| \le \sqrt{1-\cos^2{{(\theta_i^a)}^*_{\pm}}} \Big\{\frac{3 \sigma r \pm \sqrt{\sigma^2 r^2 +32 \lambda^2 A^2}}{4}\Big \} 
	\label{omega_constrain_to_synch_2}
	\end{equation} 
	RHS in Eq.~\ref{omega_constrain_to_synch_2} is larger if we consider $\cos{(\theta_i^a)}^*_{+}$ in the RHS We, therefore, proceed with this choice to consider all possible oscillators into the locked state.
	In the infinite size limit, the contribution of the drifting oscillators to $r$ is $0$, which implies that the second summation in the RHS can be neglected for a large network, and therefore $r^a\simeq r^a_{locked}$. 
	In the infinite size limit, the probability of finding an oscillator in layer $a$ at phase $\theta^a$, while its mirror node's phase is $\theta^b$, is $C (\omega^a)/ |\dot \theta^a|$, where is $C (\omega^a)$ is a constant \cite{Pikosvsky2003}.
	Thus, the second term in RHS can be written as 
	\begin{equation}
	\begin{aligned}
	r^a_{drift} = \int_{-\pi}^{\pi} \int_{-\pi}^{\pi} \int_{-0.5}^{0.5} d \omega^a d \theta^a d \theta^b \times\\
	\frac{C(\omega^a) e^{i (\theta^a -\psi^a)} g(\omega^a)}{|\omega^a + \sigma r \sin(\psi^a -\theta^a) + \lambda A \; \sin(\theta^b -\theta^a)|}. \\
	\end{aligned} 
	\label{}
	\end{equation}
	Since integration over $\omega^a$ can be split up into two parts: $-0.5$ to $-\gamma_c$ and $\gamma_c$ to $0.5$, where $\gamma_c$ is equal to RHS of Eq. \ref{omega_constrain_to_synch_2}. For a symmetric natural frequency distribution,  $g(-\omega^a) = g(\omega^a)$, we get $r^a_{drift} = 0$ from the same arguments as shown in \cite{Pikosvsky2003}.
	For a given $\sigma$ and $\lambda$, neglecting the contribution of the drifting oscillators to $r^a$, we find the roots ($r^a$ values) of Eq. \ref{r_lay_a} numerically. Note that for a symmetric natural frequency distribution, one can use either $x_{i_2}$ or $x_{i_4}$ for $\pm \omega_i^a$ in Eq. \ref{r_lay_a}. 
	Fig.~\ref{analytic_numeric}(a, b) demonstrates that the analytical prediction of $r^a$ is in a fair agreement with its numerical estimation. In Fig.~\ref{analytic_numeric}(b), solid and dashed lines correspond to $x_{i_{2(4)}}$, while the line with stars corresponds to $x_{i_{1(3)}}$. If there exist two $r$ values for a given $\sigma$ and $\lambda$, the line with stars shows only the largest of them.
	It is obvious that only $x_{i_{2(4)}}$ represents the synchronous state. 
	Fig.~\ref{analytic_numeric}(a) depicts that in the absence of adaptive multiplexing, $r^a$ gradually increases yielding a partially synchronized state, a single cluster of all the nodes having $\dot \theta_i^a = 0$, which grows in size with recruiting more and more nodes in it as $\sigma$ increases. 
	In the presence of adaptive coupling, Eq.~7 does not have
any non-zero solution for $\sigma < \sigma_c^b$, whereas at $\sigma_c^b$ one observes an abrupt transition to $r^a \approx 1$ (see Fig.~\ref{analytic_numeric}(b)). In Fig.~\ref{analytic_numeric}(a), the difference between the numerics and the analytical solution corresponding to $\sigma \approx 0.4$ is approximately $0.1$, which might be arising due to the omission of the drifting oscillators in Eq. \ref{r_lay_a}. However, for smaller and larger values of $\sigma$, the difference is negligible. The solid line in Fig.~\ref{analytic_numeric}(b) refers to a stable synchronized state, and the dashed line joining the stable state to the incoherent state, refers to an unstable state \cite{Zhang2015, Zhang2013, Vladimir2015}. For any given value of $\sigma$, $r^a = r=0$ is also a solution of Eq. \ref{r_lay_a} although not shown in Fig. \ref{analytic_numeric}(b).
	At $r=a=0$, RHS of Eq. \ref{omega_constrain_to_synch_2} is zero, hence $r=0$ is a solution of Eq. \ref{r_lay_a}. 
	The presence of the unstable state is an indicator of simultaneous presence of two local attractors, i.e., incoherent state and the coherent state \cite{Leyva2013b, Zhang2015}. 
	The values obtained for $\sigma_c^f$ from numerical simulations presented in Fig.~\ref{multiplexing_effect} (b), Fig.~\ref{Hysterisis_size} and Fig.~\ref{analytic_numeric}(b) can also be perceived from \cite{Strogatz1991},
	which shows that the incoherent state loses its stability at $\sigma_c^f = 4\gamma/\pi=0.636$ in the thermodynamic limit.
	\begin{figure}[t]
		\centering
		\centerline{\includegraphics[width=\columnwidth]{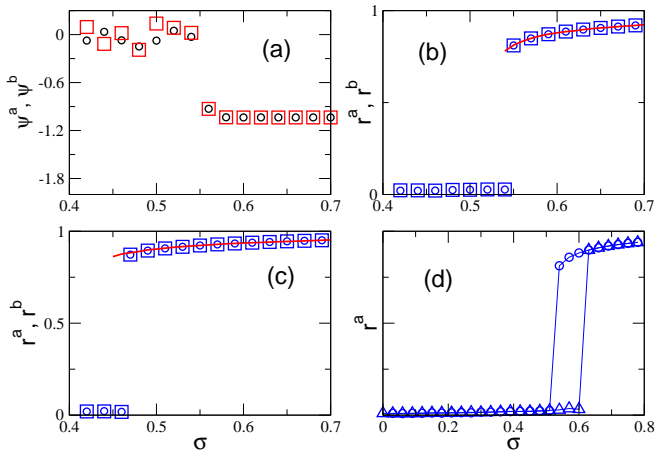}}
		\caption{(color online). (a) Time averaged $\psi^a$ (circle) and $\psi^b$ (square) in backward continuation of $\sigma$. (b, c) $r^a$ (circle), $r^b$ (square) in backward continuation of $\sigma$. (d) $r^a$ vs $\sigma$ when adaptive coupling $\lambda r^a r^b$ is replaced by $\lambda r^a$ in Eq. \ref{model} (a) and by $\lambda r^b$ in Eq. \ref{model} (b). 
			Here $\Delta \omega = 0.381$ (a, b), $0.781$ (c), $0.61$ (d). $\lambda = 0.2, N=10000$. 
			In (b, c) the continuous line corresponds to solution of Eq. \ref{r_lay_a}. Triangle, circle in (d) corresponds to forward, backward continuation of $\sigma$.
		} 
		\label{analytic_numeric_3_order_pol}
	\end{figure} 
	
	We find that adaptive coupling suppresses the formation of the giant cluster, which in turn, leads to ES. To demonstrate the suppression, we evaluate RHS in Eq. \ref{omega_constrain_to_synch_2} for the two cases, $A=1$ and $A=r^2$. For the forward continuation of $\sigma$, we start with $r \approx 0$, and therefore we can safely consider $r \approx 0$. Further to simplify the calculations we take $\sigma =\lambda$. For $A=1$ and $r^2$, $\cos{(\theta_i^a)}^*_{+} \approx 0.7$ and $0$, and hence $\sin{(\theta_i^a)}^*_{+} \approx 0.7$ and $1$, respectively, which are almost the same in both cases. 
However, the bracketed term in the RHS for the two cases is approximately $\sqrt{2}\lambda,0$, and therefore, the adaptive coupling suppresses the number of nodes in the partial synchronous state or in the other words, formation of the giant cluster is suppressed until $\sigma_c^f$ is reached.  
Furthermore, Eq.~\ref{omega_constrain_to_synch_2} exhibits that an increase in $\lambda$ leads to 
an increase in the RHS of the Eq.~\ref{omega_constrain_to_synch_2}, indicating that the same 
number of nodes can get synchronized even at a smaller $\sigma$ value, which in turn causes a 
decrease in $\sigma_c^b$.     
	For both the layers being identical, a decrease in the hysteresis width with a decrease in $\Delta\omega$ can be understood from the synchronous state. Substituting $\dot \theta_i^a = \dot \theta_i^b =0$ in Eq.~\ref{model} yields $\Delta\omega=0$, $\theta_i^a = \theta_i^b$,
leading to the cancellation of the inter-layer coupling terms and the two layers become isolated networks, which, in the thermodynamic limit manifests a discontinuous phase transition without hysteresis \cite{Pazo2005}. 
	
\subparagraph*{Mean field analysis when $\Delta \omega \neq 1$:}
	Using the mean field analysis, we prove that there exists a 
decrease in $\sigma_c^b$ with an increase in $\Delta \omega$, as well as
there exists a discontinuity in $r^a$ for globally connected layers.
	Except at the higher $\lambda$ values, in the synchronized state $r^a \gtrapprox 0.8$ 
(Fig.~\ref{Hysterisis_size}(a-c)), indicating existence of either a global synchronized state or 
almost absence of drifting oscillators. The phases in this state can be written from Eq.~\ref{root_third_order_pol} (APPENDIX A). We solve Eq.~\ref{r_lay_a} numerically by substituting $\sin(\psi^a -\theta_i^a)$ values from Eq. \ref{root_third_order_pol}. While solving Eq. \ref{r_lay_a}, only real terms, arising due to the locked oscillators, are considered in the summation.  
Fig.~\ref{analytic_numeric_3_order_pol} reflects that the numerical and the analytic prediction 
match very well. Figs.~\ref{analytic_numeric_3_order_pol}(b),(c) show that $\sigma_c^b$ decreases with an increase in $\Delta \omega$. Furthermore, Eq.~\ref{r_lay_a} does not has any non-zero 
solution below $\sigma =\sigma_c^b$ $\approx 0.45$, (Fig. \ref{analytic_numeric_3_order_pol}(c)), 
indicating absence of oscillators having $\dot \theta_i^a = \dot \theta_i^b =0$.  
	The minimum value of $r^a$ we get is approximately $0.8$, indicating suppression of the giant cluster due to the adaptive coupling.  For $\sigma > \sigma_c^b$, from the two solutions of Eq. \ref{r_lay_a}, only the largest $r^a$ value is plotted in Fig.~\ref{analytic_numeric_3_order_pol} (b),(c) i.e. we ignore the unstable solution depicted in Fig. \ref{analytic_numeric} (b). 
	Note that in the synchronous state, 
	we consider $\dot \theta_i^a = \dot \theta_i^b =0$. However, it may be possible that
only one of them is zero and other one takes a non-zero value ($\dot \theta_i^a =0$ and 
$\dot \theta_i^b \neq 0$ or vice-versa), therefore, contributions from 
these oscillators have been neglected in the summation in Eq.\ref{r_lay_a}.  
	For $\sigma < \sigma_c^b$, $r^a \approx 0$ in the numerical simulations suggests negligible
number of locked oscillators, while for $\sigma > \sigma_c^b$, almost all the oscillators are in
the locked state. Therefore avoiding the possibility of $\dot \theta_i^a =0$ and $\dot \theta_i^b \neq 0$ is a valid assumption. 
	
	{\bf Robustness of ES against change in adaptive scheme:} 
Finally, we demonstrate that ES exists even if we replace the inter-layer multiplication 
factor $r^a r^b$ in Eq.\eqref{eq1} and Eq.\eqref{eq2} by $r^a$ and $r^b$, respectively (Fig.~\ref{analytic_numeric_3_order_pol}(d)). In the infinite size limit, the addition of coupling term proportional to $r^2$ does not yield any change in the critical coupling at which incoherent state becomes unstable \cite{Strogatz1991}. Therefore, $r^a r^b$ term in the inter-layer coupling only helps 
us in fixing $\sigma_c^f$ at a constant value, which, otherwise might be sensitive to the parameters $\lambda$ or $\Delta \omega$.
	Note that in the Zhang {\em et al.} model, intra-layer coupling term contains $r^2$, which was shown to cause ES in adaptively coupled networks \cite{Zhang2015}, a condition which is not 
required for the case of adaptive inter-layer coupling.

	\section*{Conclusion}
	It is known that in a system of networked oscillators {whatever microscopic strategy} can suppress the synchronization, gives rise to ES. It was earlier reported that a fraction of adaptively intra-linked nodes through local order parameters brought about ES in a multilayer network with dependency links \cite{Zhang2015}. In the current work, an adaptive inter-linked set up between the pairs of the mirror (interconnected) nodes by means of global order parameter trigger ES in the multiplexed layers. 
	We have discussed in detail how the inter-layer coupling strength, as well as the inter-layer pairwise frequency correlation, enable us to shape the hysteresis of the emergent ES. We provide mean-field analytical treatment to ground the perceived outcomes and substantiate that the analytical predictions are in a fair agreement with the numerical estimations. 
	
	Our model has some level of similarity with several real-world complex systems having
multilayer underlying network structure. For example,
large-scale brain multilayer networks can be defined based on the functional interdependence of the brain regions \cite{brain_multilayer}. 
We hope that our investigation of ES originating from the inter-layer adaptive coupling between layers of a multiplex network would help in advance the understanding of microscopic as well as macroscopic dynamics of adaptive mechanism taking place between intertwined real-world dynamical processes.
	
	\section*{APPENDIX A}
	In general case of $\Delta \omega \neq 1$, one obtains the synchronized layers in which pairs of the mirror nodes are mutually synchronized. For globally connected layers, rewriting Eq. \ref{model} in terms of order parameter (Eq.\ref{eq_r_a}): 
	\begin{subequations}
		\begin{align}
		\dot{\theta^{a}_i} = \omega_i^{a} + \sigma r^a \sin(\psi^a-\theta_i^{a})+
		\lambda a \sin(\theta_i^{b}-\theta_i^{a}),\label{eq11a} \\
		\dot{\theta^{b}_i} = \omega_i^{b} + \sigma r^b \sin(\psi^b-\theta_i^{b})+
		\lambda a \sin(\theta_i^{a}-\theta_i^{b}),\label{eq11b}
		\end{align}\label{}
	\end{subequations}
	where $i=1,\hdots,N$. In the synchronous state, we have $\dot \theta_i^{a(b)} = \bar \omega$ \cite{Zhang2015}, where $\bar \omega$ is the mean of the natural frequencies of the entire multiplex network.
	Furthermore, considering $\bar \omega =0$ yields synchronous states with a fixed point solution.
	Numerical simulations suggest that one can assume $\psi^a \approx \psi^b =\psi$ and $r^a \approx r^b = r$ in the synchronous state (Fig. \ref{analytic_numeric_3_order_pol} (a-c)). While summing up Eq.\eqref{eq11a} and Eq.\eqref{eq11b} taking into account these approximations, yields
	\begin{equation}
	\sigma r \sin(\psi -\theta_i^b) = -\omega_i^a-\omega_i^b -\sigma r \sin(\psi -\theta_i^a)
	\label{sum_of_velocity}
	\end{equation}
	Substituting Eq.\eqref{sum_of_velocity} in Eq.\eqref{eq11b} with inter-layer coupling written as
	$\sin(\theta_i^a - \theta_i^b) = \sin(\theta_i^a - \psi)\cos(\psi - \theta_i^b) +\cos(\theta_i^a - \psi)\sin(\psi - \theta_i^b)$, where $\cos(\psi -\theta_i^{a(b)}) \approx \pm \{1-\sin^2(\psi -\theta_i^{a(b)})/2\}$. 
	Note that higher order terms in $\cos(\psi -\theta_i^{a(b)})$ are neglected as $r^{a(b)} \approx 1$ which implies that $\theta_i^{a(b)} \approx \psi^{a(b)} = \psi$ (Eq. \ref{r_lay_a}). Moreover, it also indicates that only the positive value of $\cos(\psi -\theta_i^{a(b)})$ must be considered. After some mathematical simplifications, Eq.\eqref{eq11b} results in a third order polynomial $x_i^3 + p_ix_i^2+q_ix_i+r_i' =0$, where $x_i=\sin(\psi -\theta_i^a)$. The coefficients of the polynomial are as follows:
	\begin{equation}
	\begin{split}
	p_i =& \frac{(\omega_i^a + \omega_i^b)( 3\lambda r )} { 2 \sigma r^2 \lambda},\\
	q_i =&\frac{1}{\lambda r^2} \Big\{-\sigma r -2 \lambda r^2 + \frac{(\omega_i^a + \omega_i^b)^2 \lambda }{(2 \sigma^2)} \Big \},\\
	r_i'=& -\frac{1}{\lambda r^2}\Big \{\omega_i^a + \frac{(\omega_i^a + \omega_i^b) \lambda r} { \sigma}\Big\}.
	\end{split} 
	\label{}
	\end{equation}
	
	The roots of the polynomial are given by \cite{roots}:
	\begin{equation}
	\begin{split}
	x_{i_1}=& U_i+ V_i -\frac{p_i}{3},\\
	x_{i_2}=& -\frac{(U_i+V_i)}{2} + i\frac{\sqrt{3}}{2} \frac{(U_i-V_i)}{2} -\frac{p_i}{3}, \\ 
	x_{i_3}=& -\frac{(U_i+V_i)}{2} - i\frac{\sqrt{3}}{2} \frac{(U_i-V_i)}{2} -\frac{p_i}{3}, \\
	\end{split}
	\label{root_third_order_pol}
	\end{equation}
	where
	\begin{equation}
	\begin{split} 
	U_i=& \Big\{\frac{-b_i}{2} + \sqrt{\frac{b_i^2}{4}+ \frac{a_i^3}{27}}\Big\}^{1/3},\\
	V_i=& \Big\{\frac{-b_i}{2} - \sqrt{\frac{b_i^2}{4}+ \frac{a_i^3}{27}}\Big\}^{1/3},\\
	a_i=& (3 q_i - p_i^2)/3,\\
	b_i=& (2 p_i^3 - 9 p_i q_i + 27 r_i')/27.
	\end{split} 
	\label{}
	\end{equation}
	A physically acceptable root is selected such that when $\sigma \rightarrow \infty$, $x_i \rightarrow 0$. Moreover, when $\sigma \rightarrow \infty;\ p_i \rightarrow 0;\ q_i \rightarrow -\infty;\ r_i' \rightarrow -\omega_i^a/(\lambda r^2)$. Therefore, $ a_i \rightarrow -\infty$ and $ b_i \rightarrow -\omega_i^a/(\lambda r^2)$, in turn, $\frac{b_i^2}{4}+ \frac{a_i^3}{27} \rightarrow -\infty$. A negative value of $\frac{b_i^2}{4}+ \frac{a_i^3}{27}$ implies that $U_i,V_i$ are complex numbers \cite{roots}, \textbf{writing the roots in another form leads to}
	\begin{equation}
	\begin{split} 
	x_i=& 2 \sqrt{\frac{-a_i}{3}} \cos(\frac{\phi_i}{3}+\frac{2k\pi}{3}) -\frac{p_i}{3},\\
	\end{split} 
	\label{}
	\end{equation}
	where $k=0,1,2$, and $\phi_i = \cos^{-1}\Big(\pm \sqrt{\frac{b_i^2/4}{-a_i^3/27}}\Big)$ for $b_i \lessgtr 0$. When $\sigma\rightarrow \infty$, $\phi_i = \pi/2$, which leads to $x_{i_1} =2\sqrt{\frac{-a_i}{3}} \cos(\frac{\pi}{6}),\ x_{i_2} =2\sqrt{\frac{-a_i}{3}} \cos(\frac{5\pi}{6})$ and $x_{i_3} =2\sqrt{\frac{-a_i}{3}} \cos(\frac{3\pi}{2})$. The first two roots, $x_{i_1}$ and $x_{i_2}$, diverge while $x_{i_3} \rightarrow 0$. Note that in $x_{i_3}$, $\sqrt{\frac{-a_i}{3}}$ increases while $\cos(\frac{\phi_i}{3}+\frac{4\pi}{3})$ decreases, and since the decrement is faster (as $\cos(\phi_i)$ is proportional to $a_i^{-3/2}$) than the increment, the overall product sees the decrement. Therefore, the physically accepted root is $x_{i_3}$.


	\acknowledgments 
	SJ acknowledges the Government of India, CSIR grant 25(0293)/18/EMR-II and DST grant EMR/2016/001921 for financial support. SJ also thanks the Ministry of Education and Science of the Russian Federation (Agreement No. 074-02-2018-330) for financial support. AK and ADK acknowledge CSIR, Govt. of India for SRF and RA (through project grant 25(0293)/18/EMR-II) fellowship, respectively. 

\bibliographystyle{apsrev4-1}
\bibliography{references_multiplex_2.bib}

\end{document}